# Lattice distortion and magnetic quantum phase transition in CeFeAs$_{1-x}$P$_x$O

Clarina de la Cruz[1,2], W. Z. Hu[3], Shiliang Li[1,3], Q. Huang[4], J. W. Lynn[4], M. A. Green[4], G. F. Chen[3,5], N. L. Wang[3], H. A. Mook[2], Qimiao Si[6], and Pengcheng Dai[1,2,3]

[1] Department of Physics and Astronomy, The University of Tennessee, Knoxville, Tennessee 37996-1200, USA

[2] Neutron Scattering Science Division, Oak Ridge National Laboratory, Oak Ridge, Tennessee 37831, USA

[3] Beijing National Laboratory for Condensed Matter Physics, Institute of Physics, Chinese Academy of Sciences, Beijing 100190, China

[4] NIST Center for Neutron Research, National Institute of Standards and Technology, Gaithersburg, Maryland 20899-6012 USA

[5] Department of Physics, Renmin University of China, Beijing 100872, China

[6] Department of Physics and Astronomy, Rice University, Houston, TX 77005, USA

**The discovery of ubiquitous antiferromagnetic (AF) order in the parent compounds of iron arsenide superconductors[1-6] has renewed interests in understanding the interplay between magnetism and high-transition temperature (high-$T_c$) superconductivity in these materials[7-11]. Although superconductivity in iron arsenides arises from charge carrier doping of their semimetal parent compounds[1-3], the resulting electronic phase diagrams are dramatically materials dependent,**

ranging from first-order-like AF to superconductivity phase transition for LaFeAsO$_{1-x}$F$_x$ (ref. 12), to the gradual suppression of the AF order before superconductivity for CeFeAsO$_{1-x}$F$_x$ (ref. 13), and finally to the co-existing AF order with superconductivity in SmFeAsO$_{1-x}$F$_x$ (ref. 14). To understand how the AF ground state itself can be suppressed by quantum fluctuations, it is important to isoelectronically tune the crystal lattice structure without the influence of charge carrier doping and superconductivity[11]. Here we use neutron scattering to show that replacing the larger arsenic with smaller phosphorus in CeFeAs$_{1-x}$P$_x$O simultaneously suppresses the AF order and orthorhombic distortion near $x = 0.4$, providing evidence for a magnetic quantum critical point. Furthermore, we find that the pnictogen height in iron arsenide is an important controlling parameter for their electronic and magnetic properties, and may play an important role in electron pairing and superconductivity[15,16].

In the undoped state, the parent compounds of iron arsenide superconductors such as LaFeAsO (ref. 4) and CeFeAsO (ref. 13) have an orthorhombic lattice distortion and collinear AF structure as shown in Figs. 1a and 1b. Although electron doping via the substitution of fluorine for oxygen in LaFeAsO reduces the AF order and orthorhombic lattice distortion, superconductivity appears before their complete suppression[4,17]. Applying a magnetic field to suppress the influence of superconductivity is a common means to unmask any potential quantum critical point[18], but this approach is difficult to implement here because of the large superconducting critical field for the iron pnictides[19]. We instead achieve the same purpose by using an isoelectronic phosphorus substitution for arsenic in CeFeAsO (ref. 11), recognizing that superconductivity appears neither in CeFeAsO, an antiferromagnet with orthorhombic lattice distortion[13,20], nor in CeFePO, a paramagnet with a tetragonal structure[21]. Our systematic neutron scattering studies of the structural and magnetic phase transitions in CeFeAs$_{1-x}$P$_x$O demonstrate that the pnictogen height [the average Fe-As(P) distance] and orthorhombicity of the FeAs(P) tetrahedron critically control the iron AF ordered moment and Néel temperature (Figs. 1c-1f) of the system. These results expose a magnetic quantum critical point, thereby suggesting the importance of collective magnetic quantum fluctuations to both the electronic properties in the normal state and the origin of superconductivity[7-11].

We prepared polycrystalline samples of CeFeAs$_{1-x}$P$_x$O with $x$ = 0.05, 0.10, 0.20, 0.25, 0.30, 0.35, 0.37, 0.40, 0.43 using the method similar to CeFeAsO$_{1-x}$F$_x$ (ref. 20). The resistivity and ac susceptibility measurements carried out on these samples using a commercial physical property measurement system confirmed the absence of superconductivity above $T$ = 1.8 K. Our neutron structural Rietveld refinement data were collected on the BT-1 high resolution neutron powder diffractometer at the NIST Center for Neutron Research (NCNR), Gaithersburg, Maryland[4-6,13]. We also carried out high-flux and low resolution measurements to study the magnetic structure on the HB-1, HB-1A, and HB-3 thermal triple-axis spectrometers at the High Flux Isotope Reactor (HFIR), Oak Ridge National Laboratory[4,6].

In previous work, it was found that CeFeAsO undergoes the tetragonal (space group *P4/nmm*) to orthorhombic (space group *Cmma*) structural transition below ~158 K followed by a long range commensurate AF order with a collinear spin structure below ~135 K (Figs. 1a and 1b, ref. 13). Similar to the electron doping of CeFeAsO via fluorine substitution of oxygen, we find that phosphorus replacement of arsenic on CeFeAsO also results in decreasing structural and magnetic phase transition temperatures although here no charge carriers are doped onto the FeAs layer (Fig. 1d). Within experimental accuracy, we cannot separate the structural from the magnetic phase transition for $x \geq 0.05$. In the limit of the low temperature tetragonal structure ($a_o/b_o \rightarrow 1$) near $x \approx 0.4$, the static long range ordered moment vanishes (Fig. 1e) and the system becomes tetragonal paramagnetic like CeFePO (ref. 21). Figure 1f plots the phosphorus doping dependence of the orthorhombicity ($a_o/b_o$) and Fe ordered moment at low temperature. The smooth decrease of the ordered moment that tracks the distortion suggests the presence of a lattice distortion induced magnetic quantum phase transition through phosphorus doping.

To illustrate how the phase diagram in Fig. 1f is established, we show in Fig. 1c the phosphorus doping dependence of the nuclear $(1,1,1)_O$ and magnetic $(1,0,2)_M$ Bragg peaks (orthorhombic indexing) obtained using triple-axis spectrometers (TAS). The $(1,0,2)_M$ magnetic peak was chosen because its structure factor has no contribution from the Ce moment. Similar to CeFeAsO$_{1-x}$F$_x$ (ref. 13), we find that phosphorus substitution of arsenic only suppresses the Fe ordered moment and does not change the commensurate magnetic ordering wave vector (Fig. 1c). The rapid suppression of the $(1,0,2)_M$ magnetic scattering relative to the nuclear $(1,1,1)_O$ peak is seen near $x \approx 0.4$, suggesting the presence of a magnetic quantum phase transition.

Figure 2 summarizes the Ce and Fe Néel temperatures of CeFeAs$_{1-x}$P$_x$O obtained by measuring the temperature dependence of the Ce magnetic peak $(0,0,1)_M$ (ref. 13) and Fe $(1,0,2)_M$ magnetic scattering. Inspection of the inset of Fig. 2a reveals that the Ce long-range ordered moment decreases very rapidly with increasing phosphorus concentration and becomes difficult to detect for $x > 0.25$. The Ce ordering temperatures, however, are only weakly $x$ dependent (Fig. 2a). To quantitatively compare the phosphorus doping dependence of the Fe ordered moment and Néel temperature, we normalized the $(1,0,2)_M$ magnetic intensity to the nuclear Bragg peak and plotted their doping and temperature dependence in Figs. 2b and 2c. The Fe ordered moment and ordering temperature gradually decrease with increasing $x$, and fall below our detection limit of $\sim 0.2$ $\mu_B$/Fe for $x > 0.35$.

To see how phosphorus doping affects the tetragonal to orthorhombic lattice distortion, we plot in Fig. 3 the temperature dependence of the nuclear $(2,2,0)_T$ (here the subscript denotes tetragonal structure) peak following previous practice[13]. As the crystal structure changes from high temperature tetragonal to the low temperature orthorhombic phase, the $(2,2,0)_T$ peak will split into two peaks (see Figs. 3a-3d) and a sudden reduction in the scattering intensity at the $(2,2,0)_T$ peak position reveals the transition temperature for the lattice distortion (Fig. 3g). With increasing phosphorus concentration, the tetragonal to orthorhombic lattice distortion transition temperature decreases gradually. Although one can still observe a clear kink in temperature dependence of the $(2,2,0)_T$ peak intensity near $\sim 80$ K indicative of the tetragonal to orthorhombic phase transition for $x = 0.3$ (Fig. 3), we were unable to identify the structural phase transition temperatures (if any) for $x > 0.3$. The temperature dependence of the $(2,2,0)_T$ peak profile for x=0.35 in Fig. 3e confirms the low temperature phase to be orthorhombic while the tetragonal phase is maintained at all temperatures for x=0.40 as shown in Fig. 3f.

For a complete determination of the phosphorus doping evolution of the low temperature crystal structure, we collected full diffraction patterns at low temperatures on BT-1 (Fig. 4a). The outcome of our detailed Rietveld analysis is shown in Tables 1a and 1b. Figures 4b-4e summarize the P-doping dependence of the crystal structure in $CeFeAs_{1-x}P_xO$. The undoped CeFeAsO has an orthorhombic low-temperature structure with $c_o > a_o > b_o$ (Fig. 4b). P-doping reduces all three lattice parameters with the system becoming tetragonal for $x > 0.37$ (inset in Fig. 4b). The decreased $c$-axis lattice constant is achieved via a reduction of the As(P)-Fe-As(P) block distance [through reducing the FeAs(P) height when the larger arsenic is replaced by the smaller phosphorus], while the CeAs(P) and CeO block distances remain essentially unchanged with increasing P-doping (Fig. 4d). These trends are completely different from that of $CeFeAsO_{1-x}F_x$, where the $c$-axis lattice constant contraction is achieved via a large reduction of the Ce-As distance, while the Ce-O/F and As-Fe-As block distances actually increase with increasing F-doping. These fundamental differences reflect the fact that F-doping brings the Ce-O/F charge transfer layer closer to the superconducting As-Fe-As block and facilitates electron transfer, while P-doping is a pure geometrical lattice effect without charge carrier transfer. Furthermore, since the Fe-As distance (2.405 Å) is essentially doping independent in $CeFeAsO_{1-x}F_x$ (ref. 13) but decreases rapidly with increasing P-doping and saturates at ~2.36 Å for $CeFeAs_{1-x}P_xO$ above $x \sim 0.37$ (Fig. 4c), the reduced Fe-As(P) distance may induce strong hybridization between the Fe $3d$ and the As $4p$ orbitals and thus quench the ordered Fe magnetic moment in CeFeAsO (refs. 22-24). More generally, the P-doping increases the electronic kinetic energy thereby weakening the magnetic order[25].

Our demonstration of the pnictogen height as a parameter to systematically tune the electronic and magnetic properties of CeFeAs$_{1-x}$P$_x$O is a new insight that is likely of broad relevance to all families of the iron pnictides. In principle, the changing pnictogen height can affect both the superconducting gap function and the strength of the electron pairing[15,16]. Differences in the pnictogen heights have been used to understand the nodeless and nodal superconducting gap functions in LaFeAsO$_{1-x}$F$_x$ and LaFePO, respectively. For CeFeAs$_{1-x}$P$_x$O, the simultaneous disappearance of orthorhombicity and the static Fe ordered moment reveals a strong coupling between the collinear AF order and orthorhombic structure, similar to that in the Ba(Sr,Ca)Fe$_2$As$_2$ family of materials[5,26-28]. These results suggest that the orthorhombic lattice distortion and, potentially, orbital effects, play an important role in establishing the collinear AF order and may also be related to superconductivity of these materials[29]. Detailed theoretical studies of the role of the pnictogen height on the evolution of the magnetic structure of the phosphorus doped parent iron arsenides will clearly be very instructive.

In summary, we have mapped out the structural and magnetic phase transitions of CeFeAs$_{1-x}$P$_x$O. We showed that superconductivity does not appear in the magnetic phase diagram over the entire measured doping and temperature ranges, and found the surprising property that the lattice distortion and the static Fe AF order are gradually suppressed by isoelectronic phosphorus substitution of arsenic. These results indicate the presence of a magnetic quantum critical point near phosphorus concentration of $x \sim 0.4$, and reveal that the pnictogen height in iron arsenic tetrahedron directly controls the electronic and magnetic properties of these materials. Therefore, electron-lattice interactions are an important ingredient that must be taken into account in our understanding the electronic properties of iron arsenide superconductors.

29. Krüger, F., Kumar, S., Zannen, J. & van den Brink, J., Spin-orbital frustrations and anomalous metallic state in iron-pnictide superconductors. *Phys. Rev. B* **79**, 054504 (2009).**Acknowledgements.** We thank C. Geibel for helpful discussions. This work is supported by the US National Science Foundation through DMR-0756568, by the US Department of Energy, Division of Materials Science, Basic Energy Sciences, through DOE DE-FG02-05ER46202. This work is also supported in part by the US Department of Energy, Division of Scientific User Facilities, Basic Energy Sciences. The work at the Institute of Physics, Chinese Academy of Sciences, is supported by the National Science Foundation of China, the Chinese Academy of Sciences and the Ministry of Science and Technology of China. The work at Rice University is supported by the US National Science Foundation through DMR-0706625 and the Robert A. Welch foundation.

**Author Information** The authors declare no competing financial interests. Correspondence and requests for materials should be addressed to P.D. (daip@ornl.gov).11

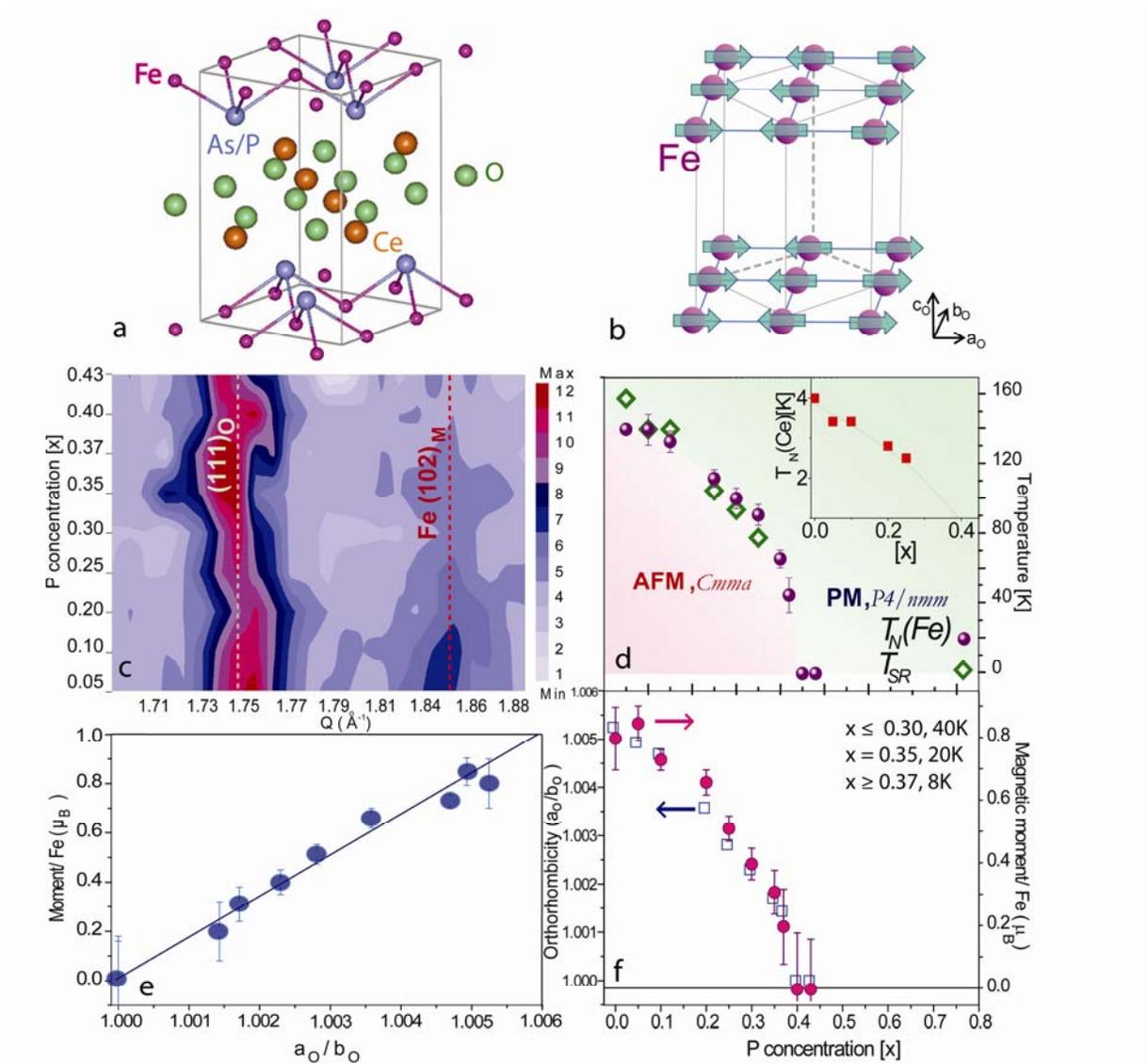

**Figure 1. Low temperature lattice and magnetic structures for Fe in CeFeAsO and the P substitution dependence of the structural/magnetic phase diagram of CeFeAs$_{1-x}$P$_x$O. a**) The three dimensional crystal lattice structure of CeFeAsO. **b**) The magnetic unit cell of Fe in CeFeAsO. The Fe moments lie in the *a-b* plane and form an antiferromagnetic collinear structure similar to that of LaFeAsO (ref. 4), while nearest-neighbor spins along the *c*-axis are parallel and so there is no need to double the magnetic cell along the *c*-axis[13]. **c**) Magnetic scattering at the Fe (1,0,2)$_M$ peak, collected using triple-axis spectrometers at ORNL, normalized to the nuclear Bragg peak (1,1,1)$_O$. The

vanishing magnetic peak intensity is clearly seen near $x \sim 0.4$. **d**) Experimentally obtained structural and magnetic phase transitions temperatures as a function of P-substitution of As. The inset in **d**) shows the P-dependence of the Ce AF ordering temperatures. **e**) The ordered AF moment of Fe is proportional to the orthorhombicity (defined as $a_o/b_o$) of the system and becomes zero at $a_o/b_o = 1$. **f**) P-doping dependence of the orthorhombicity and Fe ordered moments. The Fe ordered moments measured temperatures are noted in the inset. The error bars indicate one standard deviation.

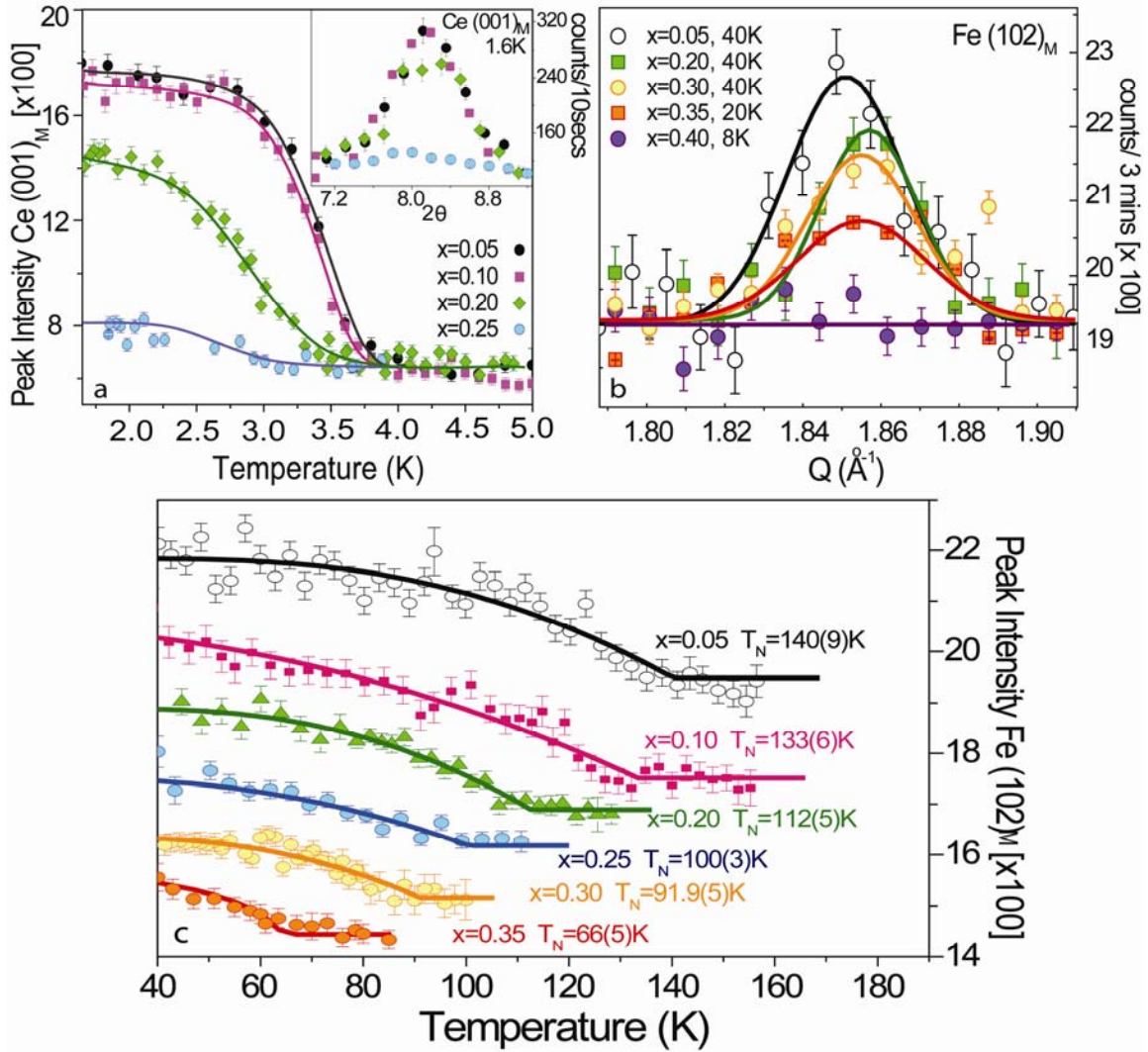

**Figure 2 Ce and Fe magnetic phase transition temperatures as a function of increasing P substitution in CeFeAs$_{1-x}$P$_x$O.** The data were collected at HFIR triple-axis spectrometers using incident beam energies of either $E_i$ = 13.5 or 14.7 meV with pyrolytic graphite as monochromator and filter. **a)** Temperature dependence of the Ce $(0,0,1)_M$ magnetic Bragg peak for various $x$. The inset shows the $(0,0,1)_M$ peak profile at 1.6 K as a function of $x$. **b)** Doping dependence of the Fe $(1,0,2)_M$ magnetic Bragg peak normalized to the nuclear $(1,1,1)_O$ peak intensity. The AF peaks appear at the commensurate ordering wave vectors $Q$ = 1.851(2), 1.850(2), 1.857(2), 1.851(2), 1.855(3), 1.854(4), 1.848(4) Å$^{-1}$ for $x$ = 0.05, 0.10, 0.20, 0.25, 0.30, 0.35, 0.37, respectively. The corresponding spin-spin correlation lengths are $\xi$ = 205(24), 179(17), 175(17), 157(23), 174(28), 140(33), 130(29) Å. The peak positions are essentially doping independent, while the spin-spin correlation length decreases with increasing P concentration. **c)** Temperature dependence of the order parameter at the Fe magnetic Bragg peak position $(1, 0, 2)_M$ as a function of P concentration. The Fe ordered moments were estimated from comparing the magnetic scattering intensity with nuclear Bragg intensity at 40 K for $x \leq 0.3$, at 20 K for $x = 0.35$, and at 8 K for $x =$ 0.37, 0.40, 0.43. The solid lines are mean field fits to the data. The error bars indicate one standard deviation.

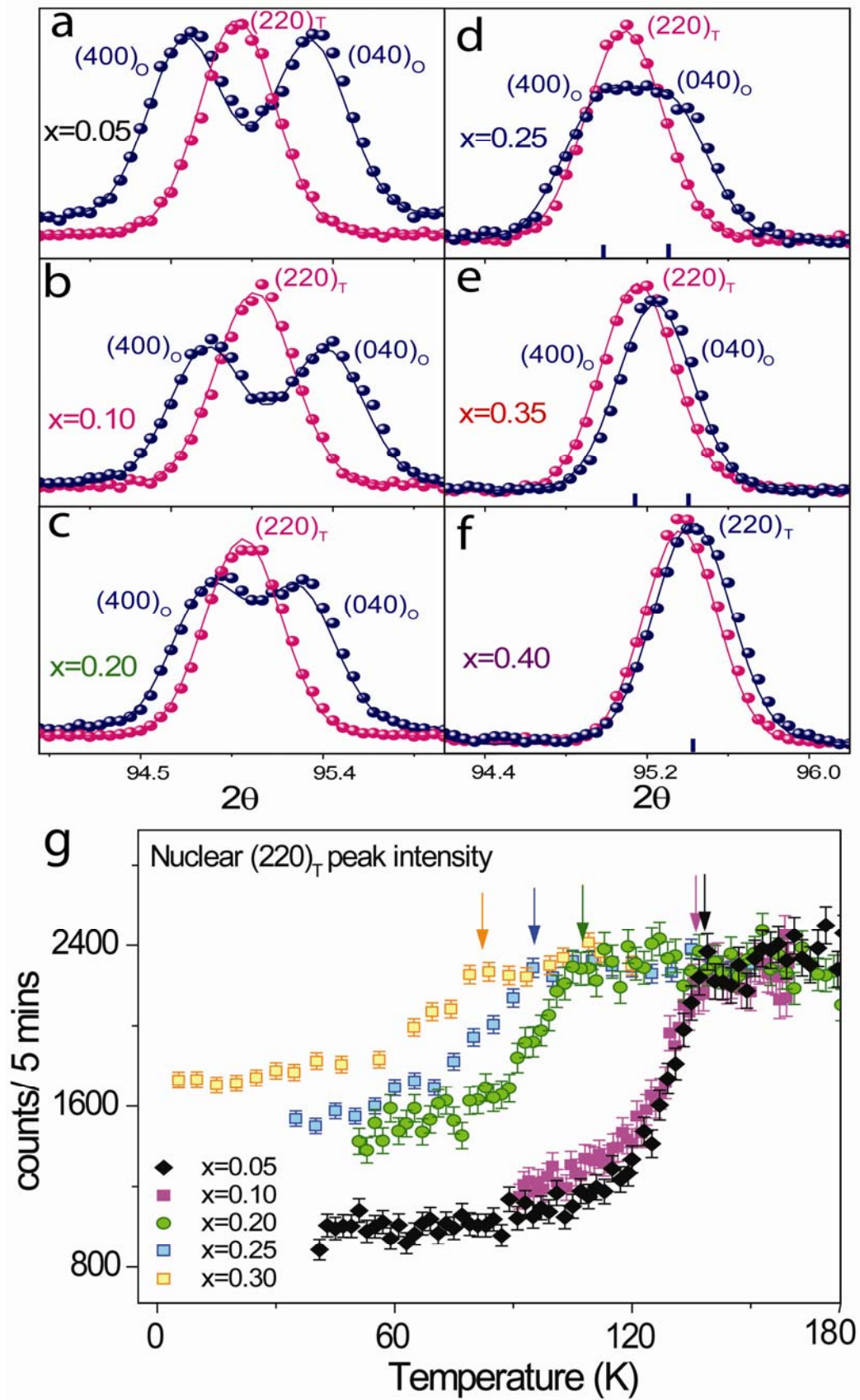

**Figure 3 Low temperature lattice structure and tetragonal to orthorhombic structural phase transition temperatures for $CeFeAs_{1-x}P_xO$.** The data were collected on BT-1 at NCNR.  **a-f**) Temperature dependence of the $(2,2,0)_T$ (T denotes tetragonal) nuclear reflection at temperatures above and below the structural phase transition for various $x$. In most cases ($x \leq 0.35$), the low temperature width is broader than that at high temperatures.   **g**) Temperature dependence of the $(2,2,0)_T$ Bragg peak intensity for different samples, which shows clear kinks indicating tetragonal to orthorhombic phase transitions.  The error bars indicate one standard deviation.

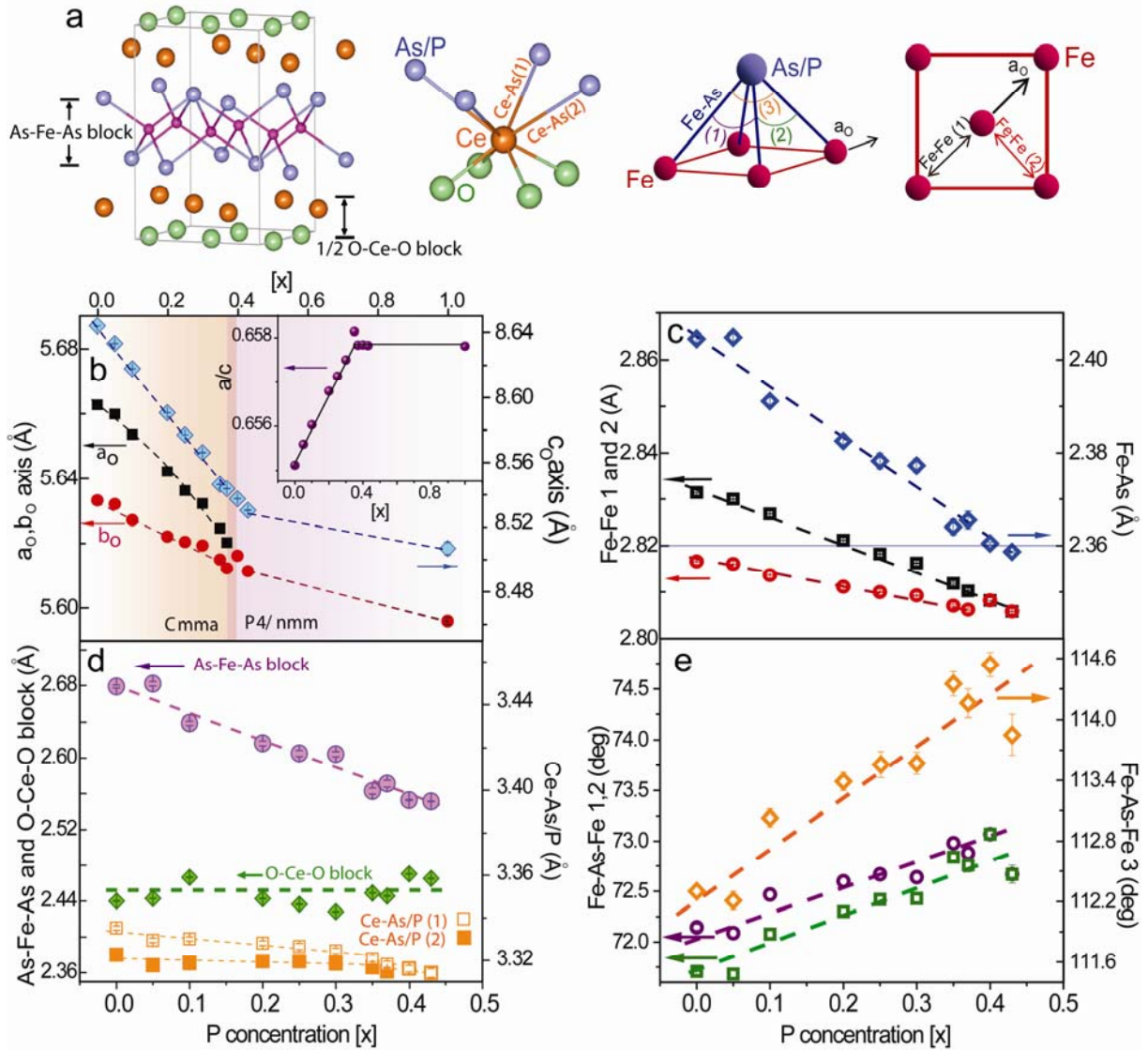

**Figure 4 Low temperature structural evolution of CeFeAs$_{1-x}$P$_x$O as a function of P concentration obtained from analysis of the BT-1 data.** The atomic positions of CeFeAs$_{1-x}$P$_x$O are shown in Tables 1a and 1b. The effect of the smaller P substitution to the larger As is to contract the Fe-As-Fe block and reduce the average Fe-As distance, thereby increasing the Fe-As hybridization. **a**) schematic diagrams defining the Fe-As-Fe and CeO blocks and illustrating various bond distances/bond angles. **b**) *a*, *b*, *c* lattice constants of the orthorhombic unit cell as a function of *x*. The inset shows that *a/c* ratio saturates when the Fe moment vanishes. **c**) Fe-Fe and Fe-As distances as a function of P

concentration. The Fe-As distance falls on the critical value of 2.36 Å where the Fe ordered moment is suppressed. **d)** Ce-O/F and Ce-As distances are essentially P concentration independent. The O-Ce-O block size is also P substitution independent. The effect of P substitution of As is to reduce the Fe-As bond distance and As-Fe-As block size. This is purely due to the size (lattice) differences between P and As. **e)** Fe-As-Fe bond angles as defined in **a)** versus P doping. The most dramatic effect of P-doping is to increase the Fe-As-Fe angle to 114.6 degrees and decrease the corresponding As-Fe-As angle to 107.3 degrees. Thus, P doping makes the FeAs tetrahedra deviate further from the ideal As-Fe-As angle of 109.5 degrees. The error bars indicate one standard deviation.

**Table 1a**. Refined structure parameters of $CeFeAs_{1-x}P_xO$ with $x$ = 0.05, 0.10, 0.20, 0.25, 0.30, 0.35 and 0.37 at 40 K for $x$ = 0.05-0.30 and 8 K for $x$ > 0.35. Space group: *Cmma* ($x$ = 0.05-0.37). Atomic positions: Ce: 4$g$ (0, ¼, $z$); Fe: 4$b$ (¼, 0, ½), As/P: 4$g$ (0, ¼, $z$), and O: 4$a$ (¼, 0, 0). The occupancies for P and As were kept at nominal values for all the refinements.

| x | a (Å) | b (Å) | c (Å) | z (Ce) | z (As) | Rp (%) | wRp (%) | $\chi^2$ |
|---|---|---|---|---|---|---|---|---|
| 0.05 | 5.65993(8) | 5.63215(8) | 8.6333(1) | 0.14149(2) | 0.6539(3) | 3.58 | 4.02 | 1.161 |
| 0.10 | 5.65362(9) | 5.62721(9) | 8.6179(1) | 0.14310(4) | 0.6529(4) | 4.18 | 4.63 | 1.614 |
| 0.20 | 5.64225(8) | 5.62212(8) | 8.5906(1) | 0.14218(4) | 0.6523(3) | 3.68 | 4.07 | 1.866 |
| 0.25 | 5.6363(1) | 5.6205(1) | 8.5771(1) | 0.14203(5) | 0.6519(4) | 4.18 | 4.34 | 1.08 |
| 0.30 | 5.6323(2) | 5.6194(2) | 8.5662(2) | 0.14173(4) | 0.6520(4) | 4.29 | 4.68 | 1.458 |
| 0.35 | 5.62465(4) | 5.61503(4) | 8.5463(2) | 0.14328(3) | 0.6504(4) | 4.66 | 5.64 | 1.869 |
| 0.37 | 5.62032(3) | 5.61229(3) | 8.54369(2) | 0.14313(5) | 0.6498(4) | 4.20 | 4.88 | 1.295 |

**Table 1c**. Refined structure parameters of CeFeAs$_{1-x}$P$_x$O with $x$ = 0.40, 0.43 at 8 K. Space group: *P4/nmm*. Atomic positions: Ce: 2*c* (¼, ¼, z); Fe: 2*b* (3/4, ¼, ½), As/P: 2*c* (¼, ¼, z), and O: 2*a* (3/4, ¼, 0).

| $x$ | $a$ (Å) | $b$ (Å) | $c$ (Å) | z (Ce) | z (As) | Rp (%) | wRp (%) | $\chi^2$ |
|---|---|---|---|---|---|---|---|---|
| 0.40 | 3.97132(4) | 5.61629(5) | 8.53749(2) | 0.14468(3) | 0.6497(3) | 4.81 | 5.83 | 2.313 |
| 0.43 | 3.96792(7) | 5.61149(7) | 8.53036(2) | 0.14449(3) | 0.6491(2) | 5.93 | 7.12 | 1.418 |